\documentclass[a4paper, 10pt,english,10pt]{scrartcl}
\usepackage{mathpazo}
\usepackage{avant}
\usepackage[T1]{fontenc}
\usepackage[latin9]{inputenc}
\usepackage{geometry}
\usepackage{color}
\usepackage{array}
\usepackage{float}
\usepackage{textcomp}
\usepackage{multirow}
\usepackage{amsmath}
\usepackage{graphicx}
\PassOptionsToPackage{normalem}{ulem}
\usepackage{ulem}

\makeatletter

\providecommand{\tabularnewline}{\\}

\newcommand{\lyxaddress}[1]{
\par {\raggedright #1
\vspace{1.4em}
\noindent\par}
}

\usepackage{subfig}

\usepackage{mathptmx}
\usepackage{courier}
\usepackage{cite}
\usepackage{numprint}

\@ifundefined{showcaptionsetup}{}{%
 \PassOptionsToPackage{caption=false}{subfig}}
\usepackage{subfig}
\makeatother

\usepackage{babel}
\begin{document}

\title{Off-axis reference beam for full-field swept-source OCT and holoscopy}

\author{Dierck Hillmann,\textsuperscript{1,*} Hendrik Spahr,\textsuperscript{3}
Helge Sudkamp,\textsuperscript{3} \\
Carola Hain,\textsuperscript{3} Laura Hinkel,\textsuperscript{1}
Gesa Franke,\textsuperscript{3}\\
Gereon Hüttmann\textsuperscript{2,3,4}}
\maketitle

\lyxaddress{\textsuperscript{1}Thorlabs GmbH, Maria-Goeppert-Str. 5, 23562 Lübeck,
Germany}

\lyxaddress{\textsuperscript{2}Medizinisches Laserzentrum Lübeck GmbH, Peter-Monnik-Weg
4, 23562 Lübeck, Germany}

\lyxaddress{\textsuperscript{3}Institut für Biomedizinische Optik, Universität
zu Lübeck, Peter-Monnik-Weg 4, 23562 Lübeck, Germany}

\lyxaddress{\textsuperscript{4}Airway Research Center North (ARCN), Member of
the German Center for Lung Research (DZL), Germany}

\lyxaddress{\textsuperscript*\textcolor{blue}{\uline{dhillmann@thorlabs.com}}}
\begin{abstract}
In numerous applications, Fourier-domain optical coherence tomography
(FD-OCT) suffers from a limited imaging depth due to signal roll-off,
a limited focal range, and autocorrelation noise. Here, we propose
a parallel full-field FD-OCT imaging method that uses a swept laser
source and an area camera in combination with an off-axis reference,
which is incident on the camera at a small angle. As in digital off-axis
holography, this angle separates autocorrelation signals and the complex
conjugated mirror image from the actual signal in Fourier space. We
demonstrate that  by reconstructing the signal term only, this approach
enables full-range imaging, i.e., it increases the imaging depth by
a factor of two, and removes autocorrelation artifacts.  The previously
demonstrated techniques of inverse scattering and holoscopy can then
numerically extend  the focal range without loss of lateral resolution
or imaging sensitivity. The resulting, significantly enhanced measurement
depth is demonstrated by imaging a porcine eye over its entire depth,
including cornea, lens, and retina. Finally, the feasibility of \textit{in
vivo} measurements is demonstrated by imaging the living human retina. 
\end{abstract}

\section{Introduction}

Fourier-domain optical coherence tomography (FD-OCT) is nowadays widely
used to image the human eye. It can acquire three-dimensional tomograms
of its anterior and posterior segment at an imaging speed and resolution
not attainable with other methods. However, standard confocal FD-OCT
relies on mechanical lateral scanning, which limits the obtainable
imaging speed. Even more importantly, when increasing the acquisition
rate, image quality suffers as the number of photons collected per
A-scan decreases \cite{Potsaid:08}. 

Therefore, lateral parallelization of the OCT data acquisition can
enhance imaging speed in two ways: First, it removes the lateral scanning
and thus the mechanical movement. Second, it illuminates all spots
simultaneously, and thus alleviates the limits implied by the maximum
permissible exposure and allows higher irradiance on the sample and
thus higher data acquisition rates without harming patients \textendash{}
particularly for retinal imaging. The highest possible parallelization
uses an area camera as detector and a tunable light source to acquire
the spectrally resolved backscattered light \cite{Povazay:06}. The
acquisition speed of this full-field swept-source OCT (FF-SS-OCT)
is finally limited by the frame rate of the camera. With frame rates
of $100,000\,\mathrm{fps}$, retina imaging at 1.5 million A-scan
per second was successfully demonstrated \cite{Bonin:10}. With even
further increased imaging speed and due to its resulting phase stability,
new applications of full-field OCT emerged recently. In particular,
it was possible to image pulse-induced pressure wave in the human
retina \cite{Spahr:15} and to computationally correct imaging aberrations
and image photo receptor cells in vivo \cite{Hillmann:2016} and even
visualize their function \cite{Hillmann15112016}.

Generally, imaging depth in FD-OCT is restricted by the spectral resolution
of the interference of sample and reference radiation. In swept-source
OCT, this is determined by the instantaneous coherence length of the
swept-source, and huge measurement depths have been demonstrated using
scanning OCT with large coherence length lasers (see e.g.~\cite{Grulkowski:12}).
On the other hand, imaging depth in FF-SS-OCT is still limited by
the instantaneous coherence length of the light source since no long-coherence
lasers are available at the required sweep rates. 

An alternative to long coherence length lasers is the use of full-range
techniques. FD-OCT is only capable to detect absolute optical path
length differences between sample and reference but not their sign.
Ultimately this ambiguity results from the impossibility to detect
the spectral phase and obtain the complex spectrum, and it reduces
the measurement depth by a factor 2. Various full-range techniques
can resolve this complex conjugate ambiguity in FD-OCT, doubling the
measurement depth. In contrast to this, in classical and digital holography
this complex conjugate ambiguity results in a twin image being overlaid
with the object information and is in most cases solved by using off-axis
reference illumination (see e.g.~\cite{LEITH:62,Schnars2002,schnars2005digital,Kim2010}).
Similarly, an off-axis reference beam in FD-OCT can also resolve this
ambiguity and was successfully applied to FF-SS-OCT for full-range
imaging \cite{doi:10.1117/12.2006436} and later to line-field OCT
\cite{Fechtig:14}.

Even when high spectral resolution (high coherence length lasers)
or full-range techniques provide large imaging depth in FD-OCT, an
additional problem arises as the Rayleigh range should be equally
large to cover this measurement depth. Outside this focal range the
images are defocussed and reduced sensitivity due to confocal gating.
Especially at high NA, where good lateral resolution is anticipated,
this becomes increasingly challenging; in particular, the imaging
depth is not sufficient to image both cornea and lens or even the
retina, and axial scanning needs to be employed (see e.g.~\cite{Yasuno:06}).
For full-field swept-source OCT, techniques of inverse scattering
\cite{Marks:07,s8063903} and holoscopy \cite{Hillmann:11,Hillmann2:12,franke:821324,10.1117/12.889485}
were shown theoretically and experimentally, respectively. These techniques
increase the focal depth by numerically compensating the defocus.

Here we demonstrate that in addition to doubling the effective imaging
depth, using the off-axis reference removes major coherent artifacts
of FF-SS-OCT and holoscopy, which are caused by self-interference
of the sample radiation. The improvement of image quality by suppressing
autocorrelation noise is investigated for imaging human retina. This
way, the SNR improves by $6\,\mathrm{dB}$, and high quality retina
images were obtained with 7.2~million A-scans per second. We also
demonstrate that the off-axis recording geometry can be used for holoscopic
imaging of an entire porcine eye, while nearly preserving both sensitivity
and lateral resolution over an imaging depth of more than $25\,\mathrm{mm}$.
Retina as well as cornea and lens were simultaneously imaged sharply. 

\section{Theory\label{subsec:OffAxisInterference}}

\begin{figure}[h]
\begin{centering}
\subfloat[]{\raggedright{}\includegraphics[width=0.45\textwidth]{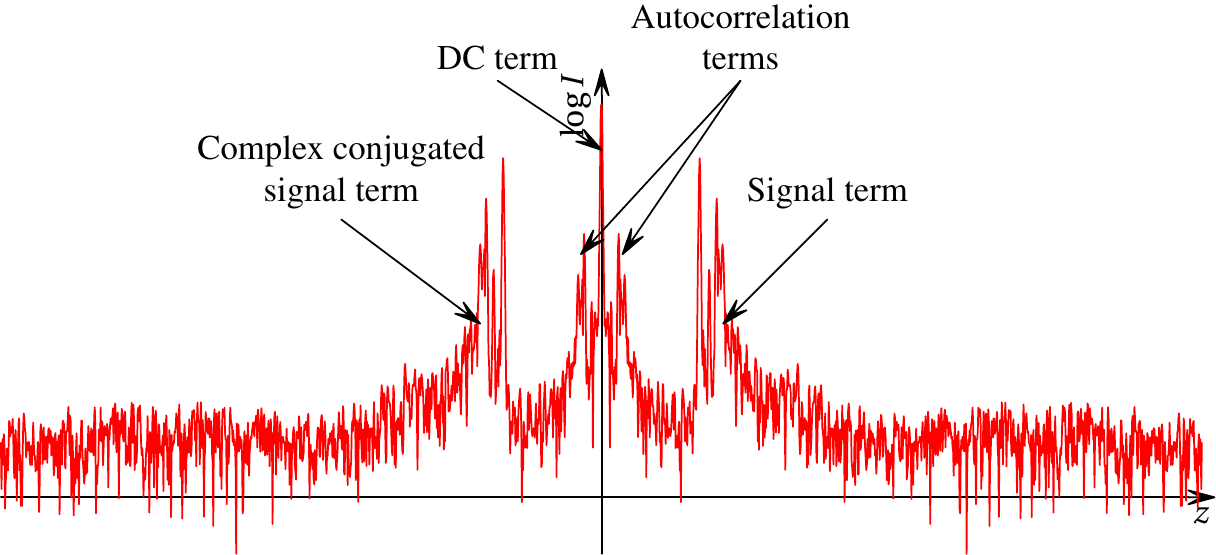}}\quad{}\subfloat[\label{fig:ImageOnCamera}]{\includegraphics[width=0.45\textwidth]{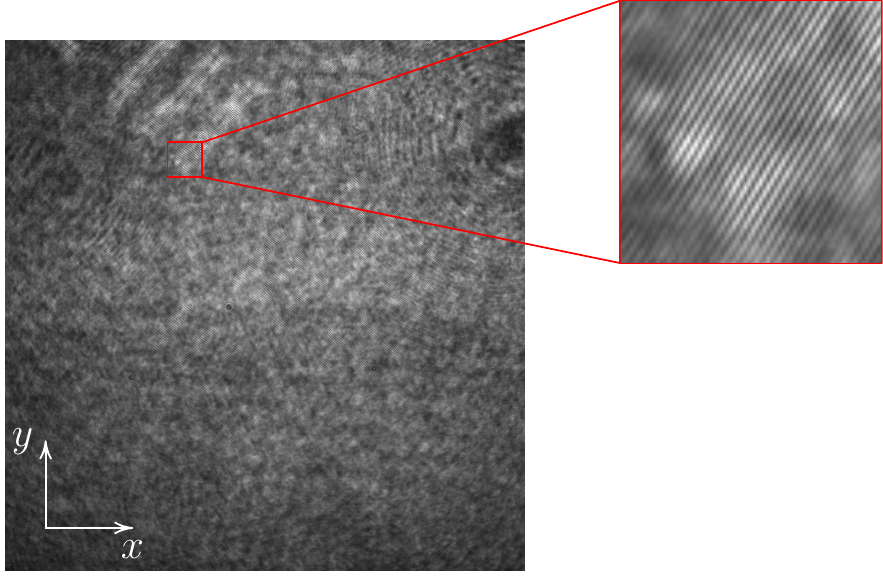}}
\par\end{centering}
\centering{}\subfloat[\label{fig:TermsDigitalHolography}]{\includegraphics[width=0.45\textwidth]{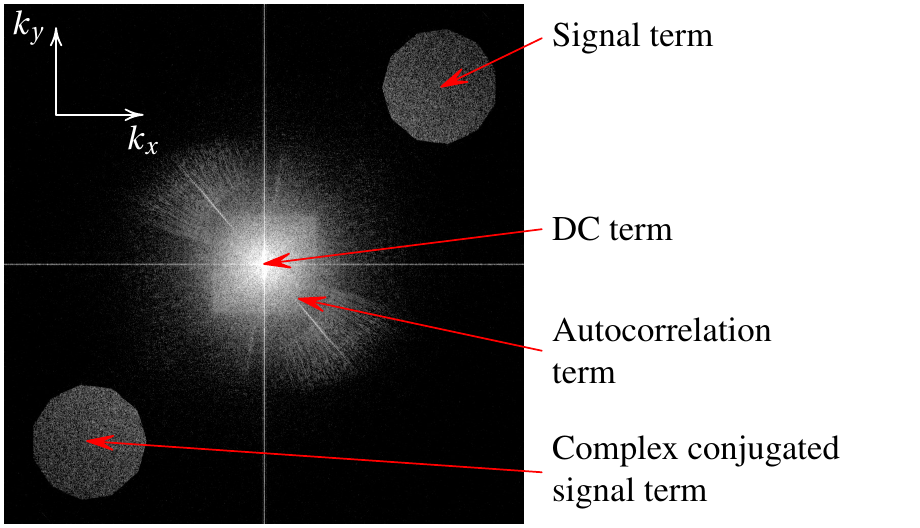}}\quad{}\subfloat[\label{fig:TermsFFSSOCT}]{\includegraphics[width=0.45\textwidth]{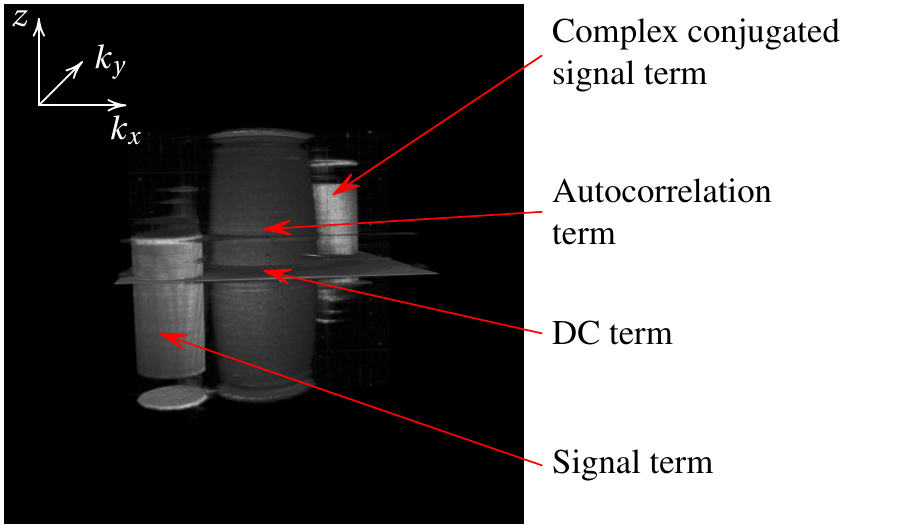}}\caption{DC, autocorrelation, and cross-correlation (signal and conjugate signal)
terms in Fourier-domain OCT. a)~A-scan of an infrared viewing card.
b)~Fringe pattern in the intensity pattern on the camera introduced
by the off-axis reference beam. c)~Fourier transform of the interference
pattern. The aperture is directly visible in the Fourier transform.
d)~Three-dimensional Fourier transform of the data cube obtained
by acquiring the interference pattern of scotch tape placed in the
image plane at 1024 equispaced wavenumbers, as recorded for FF-SS-OCT.
The spatial frequencies are shown at their correct depth. All four
terms are visible and can be separated by choosing the adequate region
in Fourier space. \label{fig:TermsOCT}}
\end{figure}

Holography, digital holography, holoscopy, and full-field swept-source
optical coherence tomography record interference images that are a
superposition of a reference wave $R\left(x,y,k\right)$ with waves
$O\left(x,y,k\right)$ scattered from a sample. Whereas in holography
only a single wavelength is employed, the latter two techniques acquire
the resulting spatial interference patterns $I\left(x,y,k\right)$
for multiple wave numbers $k$ as a function of lateral coordinates
$x$ and $y$. The recorded interference signal is given by

\begin{align*}
I\left(x,y,k\right)\propto & \left|R\left(x,y,k\right)+O\left(x,y,k\right)\right|^{2}\\
= & \left|R\right|^{2}\left(x,y,k\right)\tag{DC}\\
 & \quad+\left|O\right|^{2}\left(x,y,k\right)\mbox{\tag{autocorrelation}}\\
 & \quad+\left(R^{*}O\right)\left(x,y,k\right)\tag{signal}\\
 & \quad+\left(O^{*}R\right)\left(x,y,k\right)\mbox{\tag{conjugated signal}}.
\end{align*}
The tomographic image information is contained in the signal term
$\left(R^{*}O\right)\left(x,y,k\right)$, while the DC part, autocorrelation
signal $\left|O\right|^{2}\left(x,y,k\right)$ , and conjugated signal
$\left(O^{*}R\right)\left(x,y,k\right)$ disturb the actual imaging,
as shown for FD-OCT in Fig.~\ref{fig:TermsOCT}a. DC and autocorrelation
part are thereby caused by self-interference of reference and object,
respectively. When evaluating these data, the DC part causes a high
signal at zero-delay, the autocorrelation term causes artifacts also
referred to as coherence noise, and the conjugated signal term results
in an image mirrored at the zero-delay plane. The conjugated signal
thus results in the inability of OCT to distinguish positive time-delays
from negative time-delays and effectively reduces the imaging depth
by a factor of 2. To circumvent this limitation, full-range techniques
are capable to extract the signal term and to double the measurement
depth. However, they require additional information, as for instance
provided by small additional mechanical (axial) scanning or multiple
detectors (e.g.~\cite{Yasuno:06}). In holography, the image degradation
caused by the additional terms was first avoided by Leith and Upatnieks
by introducing an angle between the reference and the sample radiation
\cite{LEITH:62}. In off-axis holography, this causes the diffraction
angles to significantly differ for the signal term, the conjugated
signal term and the autocorrelation term. The same principle is commonly
applied in digital holography, where the signal term is isolated by
numerical filtering after Fourier transforming the recorded interference
pattern (Fig.~\ref{fig:TermsDigitalHolography}). 

For a mathematical description of the resulting signals, we assume
a reference wave on the camera that is traveling in arbitrary direction 

\[
R(x,y,k)=\left.R_{0}\exp\left(\mathrm{i}\vec{k}\cdot\vec{x}\right)\right|_{z=z_{0}}\mbox{,}
\]
where $R_{0}$ is the reference wave amplitude, $\vec{x}=\left(x,y,z\right)$
is a spatial position, $\vec{k}$ is the wave vector, also determining
the direction of the wave, and $z_{0}$ specifies the camera plane,
which is assumed to be parallel to the $xy$-plane. The wave vector
$\vec{k}$ can be decomposed into $\vec{k}_{\perp}$ and $\vec{k}_{\parallel}$,
the component orthogonal and parallel to the camera surface, respectively.
Accordingly, the reference wave can be rewritten as
\[
R\left(x,y,k\right)=R_{0}\exp\left(\mathrm{i}\vec{k}_{\parallel}\cdot\vec{x}\right)\left.\exp\left(\mathrm{i}\vec{k}_{\perp}\cdot\vec{x}\right)\right|_{z=z_{0}}\mbox{.}
\]

The first exponential only depends on the lateral $x$- and $y$-coordinates
and the second only on $z$ and is therefore constant in the camera
plane $z=z_{0}$. Computing the Fourier transform of the reference
light with respect to the $x$- and $y$-coordinates yields the zero
frequency

\[
\mathcal{F}_{xy}\left[\left|R\right|^{2}\right]=\mathcal{F}\left[R_{0}^{2}\right]=R_{0}^{2}\delta\left(k_{x},k_{y}\right)\mbox{.}
\]

With these results, a Fourier transform of the interference signal
gives 

\begin{eqnarray*}
\mathcal{F}_{xy}\left[I\left(x,y,k\right)\right] & = & \gamma\left(R_{0}^{2}\delta\left(k_{x},k_{y}\right)+\mathcal{F}_{xy}\left[\left|O\right|^{2}\right]+\vphantom{\left(R_{0}\delta\left(\vec{k}-\vec{k}_{\parallel}\right)*\mathcal{F}_{xy}\left[O\right]\right)+\left(R_{0}^{*}\delta\left(\vec{k}+\vec{k}_{\parallel}\right)*\mathcal{F}_{xy}\left[O^{*}\right]\right)}\right.\\
 &  & +\left.\vphantom{R_{0}^{2}\delta\left(k_{x},k_{y}\right)+\mathcal{F}_{xy}\left[\left|O\right|^{2}\right]}\left(R_{0}\delta\left(\vec{k}-\vec{k}_{\parallel}\right)*\mathcal{F}_{xy}\left[O^{*}\right]\right)+\left(R_{0}^{*}\delta\left(\vec{k}+\vec{k}_{\parallel}\right)*\mathcal{F}_{xy}\left[O\right]\right)\right)\mbox{,}
\end{eqnarray*}
with $*$ being the convolution operation. The signal term and conjugated
signal term are shifted by the carrier frequency $\pm\vec{k}_{\parallel}$
to different directions. The DC and autocorrelation signals are not
changed. If the carrier frequency is sufficiently large with respect
to the lateral frequency bandwidths of $O$ and $O^{*}$, they can
be separated completely in frequency space by appropriate filtering.
$O$ and $O^{*}$ have the same frequency bandwidth, while $\left|O\right|^{2}$
has twice their bandwidth. 

\begin{figure}[h]
\begin{centering}
\includegraphics{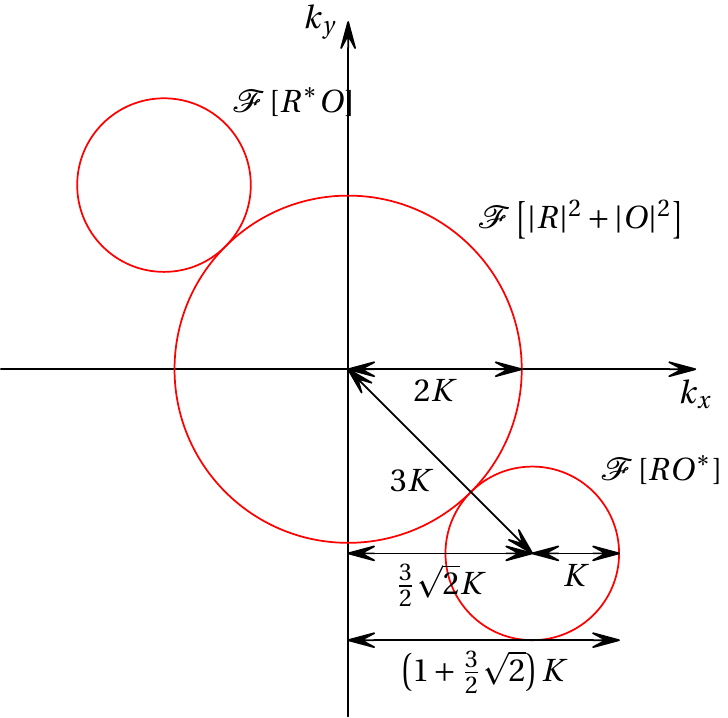}
\par\end{centering}
\caption{Reduction of the imaging bandwidth by off-axis recording. For a complete
suppression of coherent autocorrelation noise, signal terms have to
be shifted by three times their bandwidth $K$. Shifting in $x$-
or $y$-direction reduces the usable signal bandwidth by a factor
of four compared to on-axis recording. By shifting at 45\textdegree ,
usable bandwidth is only decreased by a factor of $3\sqrt{2}/2+1\approx3.1$.
\label{fig:OversamplingOffAxis}}
\end{figure}

Compared to imaging with an on-axis reference beam, the recorded lateral
bandwidth of the object field needs to be reduced when introducing
the carrier frequency, if the field of view remains the same. The
autocorrelation signal has a bandwidth twice as large as the cross-correlation
terms, and consequently the usable bandwidth $K$ is reduced by a
factor of four if the sample cross-correlation spectrum is shifted
in either $x$ or $y$ direction only. Shifting by the same amount
in $x$ and $y$ direction, the reduction in bandwidth is only a factor
of $3\sqrt{2}/2+1\approx3.1$ (Fig.~\ref{fig:OversamplingOffAxis}),
if the aperture is circular. Thus the reduction of coherent noise
comes at the cost of an approximately threefold reduced space-frequency
bandwidth product, i.e., either numerical aperture and thereby spatial
resolution or field-of-view are reduced. At the same field of view
the number of independent A-scans is reduced by the square of the
bandwidth reduction, i.e., 9.7 for a complete suppression of autocorrelation
terms; only $10\thinspace\%$ of the area in the Fourier space provided
by the detector is used for imaging. However, for practical application
efficient use of the Fourier space may be considerably increased.
However, the spectral strength of autocorrelation noise rapidly decreases
with increasing spatial frequencies. Good noise suppression is therefore
already be attained with a lower carrier frequency. Apart from this,
the usable bandwidth could additionally be increased by non-linear
filtering techniques (see e.g. \cite{Pavillon:09}). Additionally,
it should be noted hat the separation of signal and conjugate d signal
term can be achieved with a smaller off-axis angle. 

Finally, FF-SS-OCT and holoscopy acquire the interference pattern
at multiple wavelengths. However, the absolute length of the wave
vector $\vec{k}$ changes with the wavelength, while its direction
remains constant. As consequence, the off-axis terms shift in the
Fourier plane as the component parallel to the image plane, $\vec{k}_{\parallel}$,
changes as well, whereas the lateral sampling frequency of the camera
is fixed; the carrier frequencies of the signal term increases with
the wave number. This has to be taken into account when filtering
the signal term. 

\section{Materials and Methods}

\begin{figure}[h]
\begin{centering}
\includegraphics[scale=0.8]{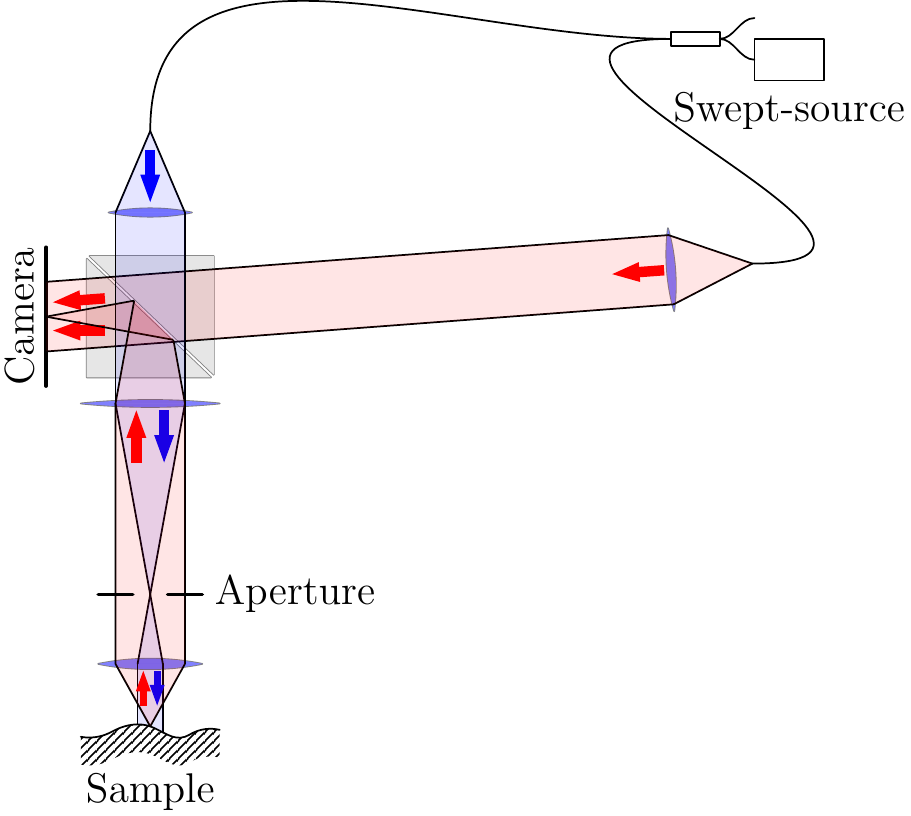}
\par\end{centering}
\caption{\label{fig:Setup}Mach-Zehnder type interferometer used for off-axis
FF-SS-OCT and holoscopy. Light emitted by a swept laser source was
split into sample and reference arm by a fiber coupler. The sample
was illuminated by a collimated beam and the backscattered light was
imaged onto the camera. The reference illuminated the camera under
a sufficient angle to separate the image from autocorrelation noise.
For \textit{in vivo} measurements the objective lens was replaced
by the lens of the eye. }
\end{figure}
A Mach-Zehnder interferometer was used for FF-SS-OCT and holoscopy.
Light from a wavelength-swept semiconductor laser (Superlum BroadSweeper
BS-840-1, $841\,\mathrm{nm}$ central wavelength, $51\,\mathrm{nm}$
tuning range, $25\,\mathrm{mW}$ output power) was split by a fiber
coupler into reference and sample arm. The sample was illuminated
with an extended parallel beam. Backscattered and reflected light
was imaged through a beam splitter cube onto a camera with suitable
magnification to ensure correct sampling. The diameter of the illuminated
sample area was matched to the imaged area. The reference light was
brought through the beam splitter cube onto the camera at a beam width
that completely illuminated the area of interest on the camera chip.
The angle of incidence of the reference light on the camera could
be adjusted, implementing both on-axis and off-axis reference illumination
for variable imaging magnifications. For off-axis imaging, the angle
was chosen to sufficiently discriminate the different signal terms
by using a live display of the Fourier transformed camera image. Polarization
of both sample and reference arm was matched to optimize contrast
in the interference signal. 

A Basler ace acA2040-180km CMOS camera was used for \textit{ex vivo}
measurements. Maximum frame rate was $180\,\mathrm{frames/s}$ at
$2048\times2048\,\mathrm{pixels}$ resolution. The wavelength sweep
of the laser was synchronized with the image acquisition to record
2048 images during one sweep.

For \textit{in vivo} measurements, a much faster CMOS camera (Photron
FASTCAM SA-Z) was used. The area of interest was reduced to $896\times368\,\mathrm{pixels}$
to enable an acquisition rate of about $60,000\,\mathrm{frames/s}$.
To further increase imaging speed, only 512 images were recorded during
one wavelength sweep. \textit{In vivo} imaging with on-axis and off-axis
reference beam resulted in different image parameters (Table~\ref{tab:EffectiveAScanRate}).
With on-axis reference imaging, the NA was up to 0.2 and an equivalent
A-Scan rate of $38.6\,\mathrm{MHz}$ was reached. For off-axis imaging
the usable bandwidth was reduced by a factor of 2.3 in $x$ and $y$
direction. To completely filter the autocorrelation noise this was
insufficient, however, it only reduced the effective A-scan rate by
a factor of 5.4. For both on-axis and off-axis imaging, the radiant
flux on the sample of $5.2\,\mathrm{mW}$ was limited by the output
of the swept source. Total energy per A-scan was $0.14\,\mathrm{nJ}$
for imaging at $0.2$ NA with the on-axis reference beam and $0.72\:\mathrm{nJ}$
with the off-axis reference beam, due to the 5.4 times reduced A-scan
rate at unchanged field of view.

\begin{table}[h]
\noindent \centering{}\caption{Imaging parameters for on-axis and off-axis FF-SS-OCT. \label{tab:EffectiveAScanRate}}
\begin{tabular}{lrr}
\hline 
 & Off-axis & On-axis\tabularnewline
\hline 
Effective A-scan rate & $7.2\,\mathrm{MHz}$ & $38.6\,\mathrm{MHz}$\tabularnewline
Numerical Aperture & 0.09 & 0.21\tabularnewline
Max. power on sample  & $5.2\:\mathrm{mW}$ & $5.2\:\mathrm{mW}$\tabularnewline
Max. power on sample per A-scan & $0.085\,\text{\textmu W}$ & $0.016\,\text{\textmu W}$\tabularnewline
Max. energy on sample per A-scan & $0.72\,\mathrm{nJ}$ & $0.14\,\mathrm{nJ}$\tabularnewline
\hline 
\end{tabular}
\end{table}

\subsection{Evaluation of signal-to-noise ratio for \textit{in vivo} measurements}

As a measure of image quality the signal-to-noise ratio of \textit{in
vivo} retinal images was determined. En-face slices of nearly homogeneous
speckle structures and of noise in front of the retina were evaluated
and for each region a histogram of the amplitude values was created.
Both, the amplitude of noise in unaveraged OCT images and of homogeneous
speckle structures in the signal follow the Rayleigh probability distribution,
as both are described by random phasor sums \cite{goodman2007speckle}.
Thus, this distribution was fit to the histograms of noise and signal,
respectively. The expectation values of the resulting distributions
were taken to compute the signal-to-noise ratio (SNR). To compare
the SNR for on-axis and off-axis, the same structures of the same
retina were evaluated. For comparison, scanned OCT images of the retina
were taken using a commercial spectrometer-based OCT system (Thorlabs
Hyperion) with a central wavelength of $840\,\mathrm{nm}$ and an
A-scan rate of $127\,\mathrm{kHz}$.

\section{Results and discussion}

\subsection{Efficient suppression of coherent noise and full-range imaging}

\begin{figure}[h]
\centering{}\subfloat[\label{fig:Tape}]{\includegraphics[width=0.22\textwidth]{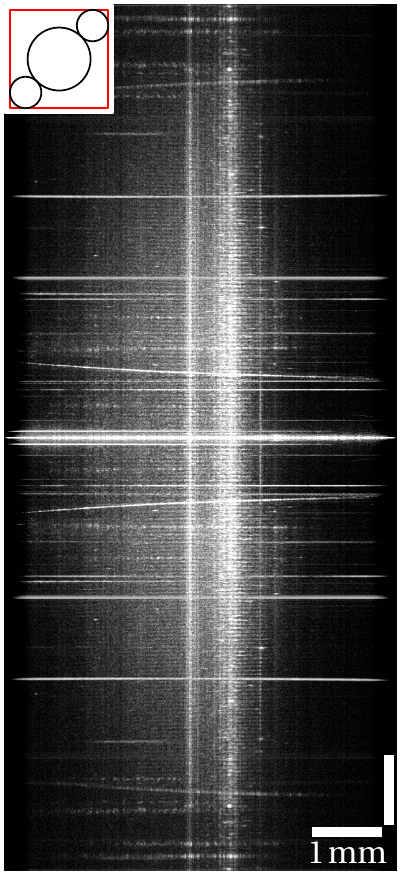}

}\quad{}\subfloat[\label{fig:TapeFiltered}]{\includegraphics[width=0.22\textwidth]{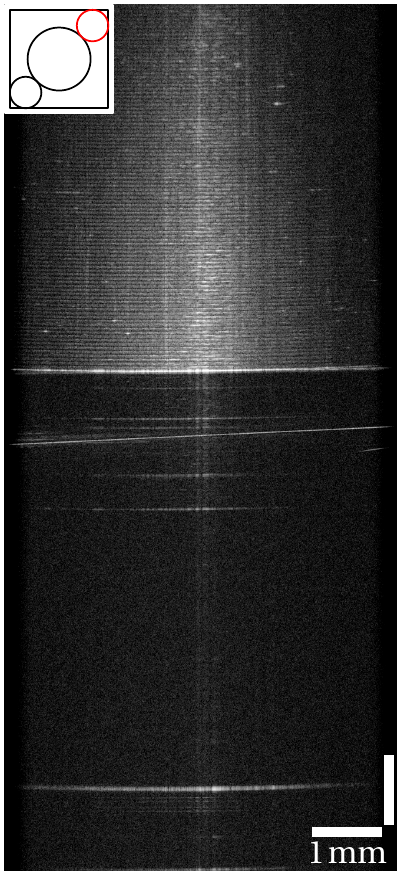}}\quad{}\subfloat[\label{fig:TapeFiltered-1}]{\includegraphics[width=0.22\textwidth]{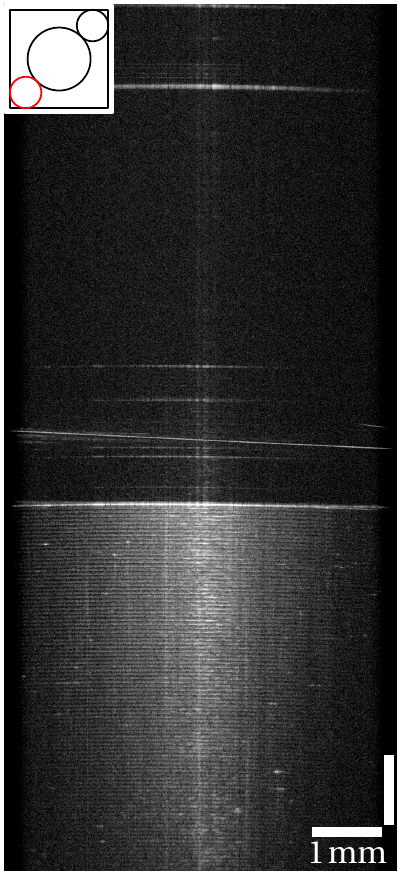}}\quad{}\subfloat[\label{fig:TapeFullRange}]{\includegraphics[width=0.22\textwidth]{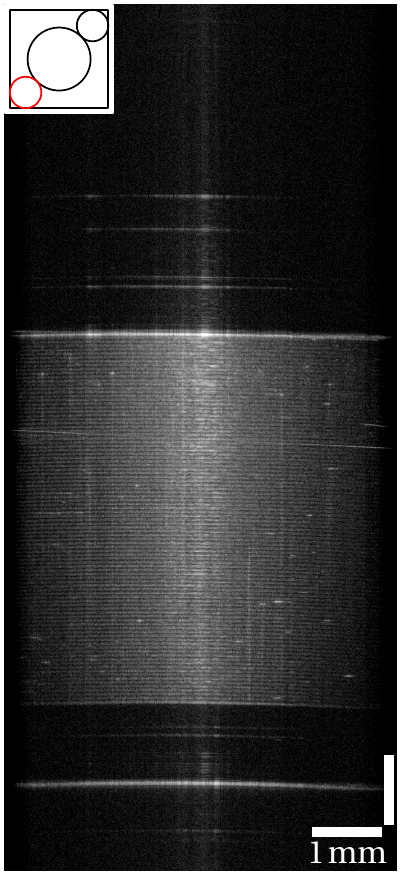}}\caption{\label{fig:ScotchTape}B-scans taken from data cubes of scotch tape
recorded with full-field swept-source OCT . a)~Without rejection
of the autocorrelation signal major artifacts are visible. Both, the
image and the conjugated image are visible. b,~c)~Reconstruction
after removal of the DC and autocorrelation terms by filtering in
the Fourier space. Coherent noise is significantly reduced. No conjugated
image is visible. The full depth range above and below the zero delay
can be used. d)~Artifact-free full-range imaging allows to position
the zero optical delay line within the scotch tape; negative and positive
path length differences are resolved. Insets show schematically filtering
(red line) in the Fourier plane.}
\end{figure}

Generally, Fourier-domain OCT images of scotch tape layers show coherent
noise. In scanning OCT this autocorrelation noise is low, since the
confocal gating suppresses out-of-focus light, and although scotch
tape layers are visible outside the Rayleigh length, they do not act
as efficient sources for interference with the other layers. On the
other hand, full-field OCT images of scotch tape layers are dominated
by coherent noise (Fig.~\ref{fig:Tape}), if they are acquired with
an on-axis reference. FF-SS-OCT and holoscopy detect reflected light
from all tape layers with the same efficiency. Thus autocorrelation
signals not only show up near the zero-delay line at the top of the
image but in the whole reconstructed volume. The strongly reflecting
layers cause autocorrelation signals significantly decreasing image
quality. 

Off-axis recording efficiently suppressed this coherent autocorrelation
noise (Fig.~\ref{fig:TapeFiltered}). When the signal term was separated
by filtering in the Fourier plane a similar image quality compared
to scanned OCT was obtained. The off-axis recording did not only separate
the image information from the autocorrelation, but it also resolved
the ambiguity with respect to positive and negative path length differences
between sample and reference radiation. When the zero path length
delay is within the scotch tape, all layers are clearly visible without
overlap at their correct position, either below or above the zero
path length (Fig.~\ref{fig:TapeFullRange}). Thus off-axis recording
of the interference pattern and filtering in the Fourier domain enabled
full-range OCT with minimal artifacts at the zero delay. 

\subsection{\textit{Improved SNR for in-vivo} measurements}

\begin{figure}[tph]
\begin{centering}
\subfloat[]{\includegraphics[width=0.375\textwidth]{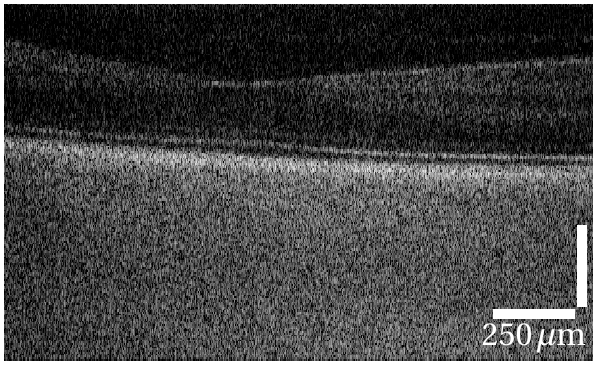}
}\quad{}\subfloat[]{\includegraphics[width=0.375\textwidth]{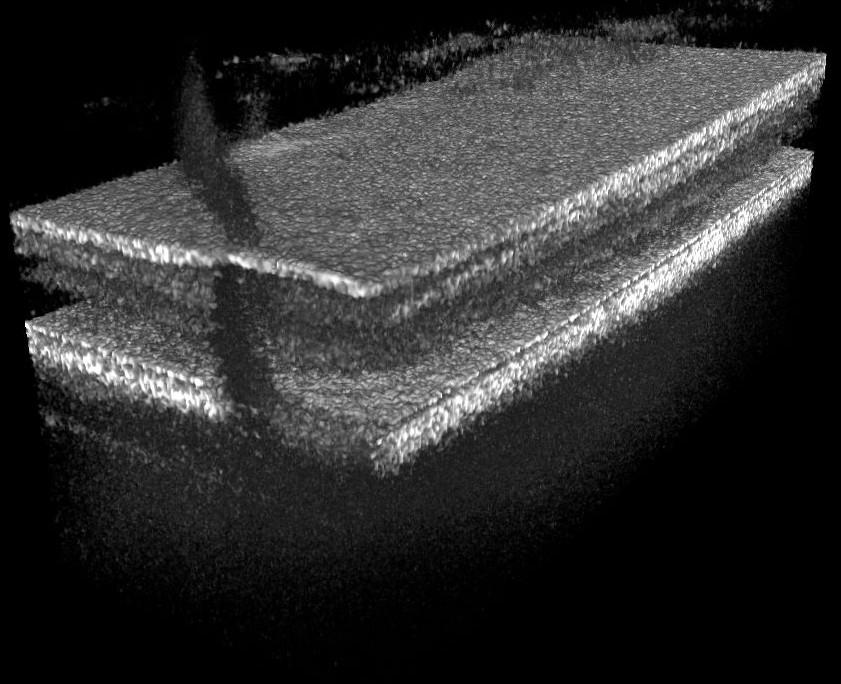}}
\par\end{centering}
\begin{centering}
\subfloat[]{\includegraphics[width=0.375\textwidth]{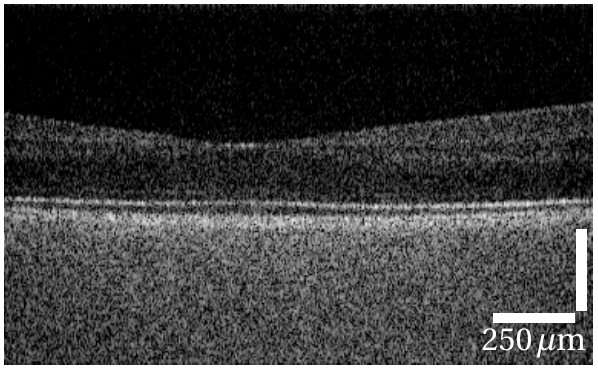}

}\quad{}\subfloat[]{\includegraphics[width=0.375\textwidth]{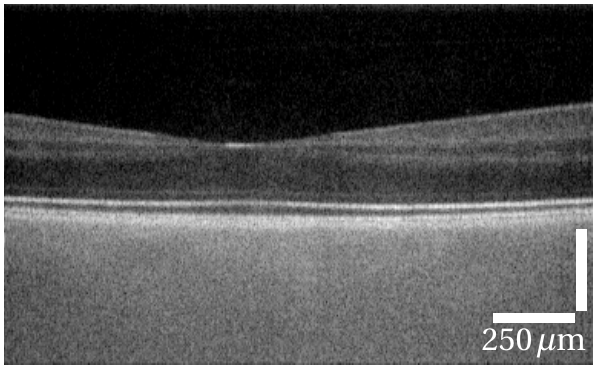}

}
\par\end{centering}
\begin{centering}
\subfloat[]{\includegraphics[width=0.375\textwidth]{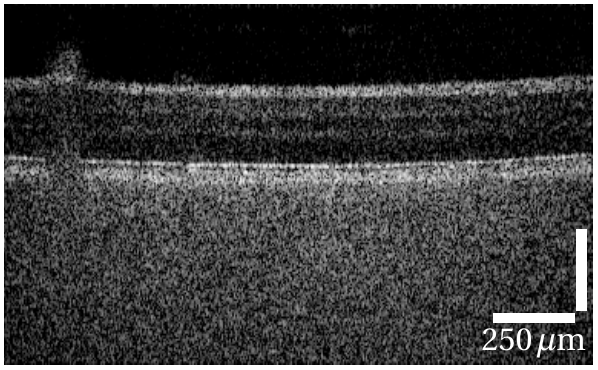}

}\quad{}\subfloat[]{\includegraphics[width=0.375\textwidth]{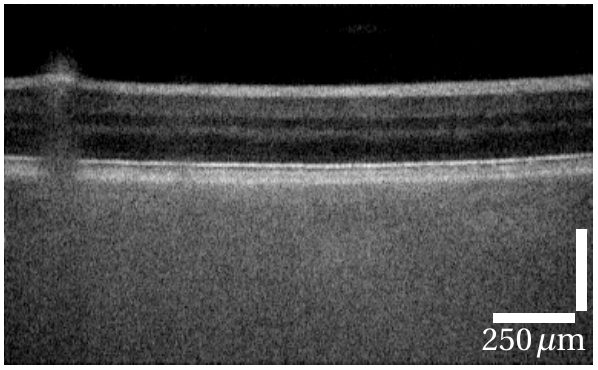}

}
\par\end{centering}
\centering{}\subfloat[\label{fig:RetinaLayers}]{\includegraphics[width=0.75\textwidth]{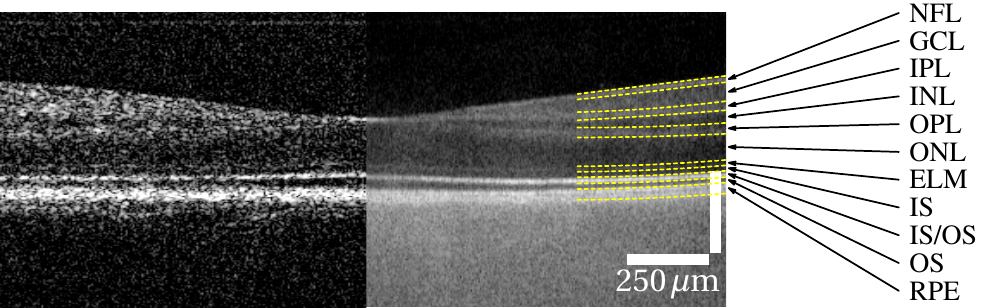}

}\caption{\textit{In vivo} retinal images acquired with OCT. a)~Macula imaged
with on-axis FF-SS-OCT. b) Volume rendering of the dataset shown in
f. c,~d)~Off-axis FF-SS-OCT images of the macula; unaveraged single
slice and average of ten lateral B-scans, respectively. e,~f)~Off-axis
FF-SS-OCT images of retinal periphery; unaveraged single slice and
average of 10 lateral B-Scans, respectively. g)~Macular region of
human retina imaged by FF-SS-OCT in comparison to scanning OCT. Layers
of the retina visible in FF-SS-OCT. NFL = nerve fiber layer, GCL =
ganglioncell layer, IPL = inner plexiform layer, INL = inner nuclear
layer, OPL = outer plexiform layer, ONL = outer nuclear layer, ELM=
external limiting membrane, IS = photoreceptor inner segments, IS/OS
= photoreceptor inner and outer segment junction, OS = photoreceptor
outer segments, RPE = retinal pigment epithelium.\label{fig:FF-RetinalImages}}
\end{figure}
Using the high speed CMOS camera human retina was measured \textit{in
vivo} at $7.2$ million A-scans/s with off-axis reference (Fig.~\ref{fig:FF-RetinalImages}a,
c, e). To enhance the imaging quality, which was compromised by the
low signal level at the high imaging speed, volumetric data sets of
OCT images were cross-correlated, shifted accordingly in all three
dimensions, and averaged (Fig.~\ref{fig:FF-RetinalImages}b, d, f).

Given the acquisition speed, which was at significalty faster than
that of commercially available OCT devices, image quality of the full-field
system is remarkable. All retinal layers that are usually visible
in OCT imaging can be clearly distinguished (see Fig.~\ref{fig:RetinaLayers}).
Intensities of the different layers above the RPE differ only slightly
between FF-SS-OCT and scanned OCT. Below the RPE the choroid shows
high signal intensities, but hardly any structures. Here mostly multiple
scattered photons were detected due to the lack of a confocal gating.
Photons from the strongly scattering RPE \cite{Baumann:12} or choroid
are always assigned to depths higher than the depth of any of their
scattering events and thus below the RPE. Therefore they disturb imaging
of the choroid, but do not corrupt imaging of the neuronal layers.
\begin{table}[H]
\noindent \begin{centering}
\captionabove{Signal-to-noise ratio (SNR) at the retinal pigment epithelium (RPE)
for different measurement geometries and sample power image at the
macula. While the SNR in on-axis geometry for $10\times$ radiant
flux on the sample increased by $1.2\,\mathrm{dB}$, the SNR can be
significantly increased by using off-axis reference illumination and
lateral filtering. \label{tab:SNRGain}}
\par\end{centering}
\noindent \centering{}%
\begin{tabular}{ccc}
\hline 
\multirow{2}{*}{Sample power} & \multicolumn{2}{c}{SNR (dB)}\tabularnewline
 & On-axis & Off-axis\tabularnewline
\hline 
$0.5\:\mathrm{W}$ & 17.7 & 17.2\tabularnewline
$5\:\mathrm{mW}$ & 18.9 & 25.2\tabularnewline
\hline 
SNR gain (dB) & 1.2 & 8.0\tabularnewline
\hline 
\end{tabular}
\end{table}

Although with off-axis imaging the lateral dimension and therefore
radiant flux and energy of each A-scan are increased, the overall
detection aperture is decreased in order to sample the interference
pattern correctly; both effects, higher flux and reduced aperture,
are expected to compensate each other if scattering is isotropic.
Hence, a similiar SNR for on-axis and off-axis imaging is expected,
if shot noise dominates coherence noise. And indeed, at 0.5 mW on
the sample similar SNR values of $17.7\,\mathrm{dB}$ and $17.2\,\mathrm{dB}$
were observed for on-axis and off-axis imaging, respectively (see
Table~\ref{tab:SNRGain}). 

When using on-axis FF-SS-OCT, a tenfold increase of the radiant power
on the sample from $0.5\:\mathrm{mW}$ to $5\,\mathrm{mW}$ improved
SNR only by $1.2\:\mathrm{dB}$ instead of the expected $10\,\mathrm{dB}$.
Since the signal is proportional to the square root of the number
of photo electrons $N_{O}$ from the sample 
\[
S\propto\sqrt{N_{R}N_{O}}\mbox{,}
\]
and the noise is dominated by shot noise 
\[
\sigma_{\gamma}=\sqrt{N_{R}+N_{O}}\mbox{,}
\]
with $N_{R}$ photo electrons detected from the reference arm, for
$N_{R}\text{\ensuremath{\ll}}N_{O}$ the SNR will increase with the
square root of $N_{O}$.

\begin{figure}[H]
\noindent \centering{}\includegraphics{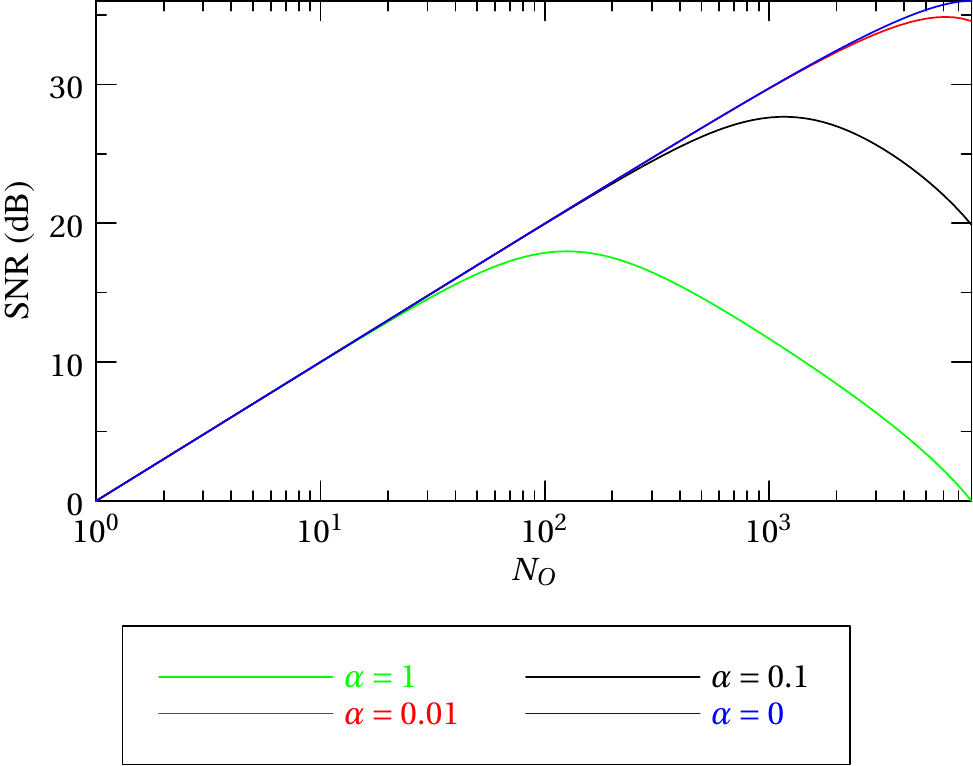}\caption{Signal-to-noise ratio for different $\alpha$ for a fixed full well
capacity of $N=\numprint{16000}$ electrons. A tenfold radiant flux
on the sample will increase the SNR by $10\:\mathrm{dB}$ for low
$N_{O}$ and low $\alpha$. For small values of $\alpha$ the SNR
increases until $N_{O}=N/2$, i.e. until reference and sample intensity
are identical. For high values of $\alpha$ the SNR decreases rapidly
when going to a significant amount of sample photo electrons. \label{fig:SNRTheory}}
\end{figure}

However, the autocorrelation term, i.e.,~coherence noise, can become
significant compared to the shot noise $\sigma_{\gamma}$, if the
number of sample photons $N_{O}$ increases in comparison to the number
of reference photons $N_{R}$. The coherence noise $\sigma_{AC}$
is directly proportional to the number of photo electrons from the
sample, i.e.,
\[
\sigma_{AC}=\alpha N_{O}\mbox{,}
\]
with $\alpha\le1$ being a suitable proportionality constant depending
on the signal depth $z$ and the sum of the autocorrelation noise
from all depths not exceeding the number of sample photons. This yields
\[
SNR\propto\frac{\sqrt{N_{R}N_{O}}}{\sqrt{N_{R}+N_{O}+\alpha^{2}N_{O}^{2}}}\mbox{.}
\]
Finally, restricting the total number of photo electrons to the full
well capacity $N$ of the detector, i.e., imposing 
\[
N=N_{R}+N_{O}\mbox{,}
\]
the SNR scales with $N_{O}$ as shown in Figure~\ref{fig:SNRTheory}
for different values of $\alpha$. 

It is clearly seen that the SNR is only expected to increase linearly
with $\sqrt{N_{O}}$ for a small number of sample photons and a low
value of $\alpha$. At higher numbers of sample photons the linear
increase of coherence noise may even decrease the SNR when illumination
of the sample is increased. Consequently, a filtering of coherence
noise, as achieved by off-axis reference illumination and suitable
filtering, should increase the SNR for high irradiation on the sample.
Indeed, as shown in Table~\ref{tab:SNRGain}, the off-axis reference
allows for an SNR gain of $8\,\mathrm{dB}$ being close to the theoretically
expected maximum of $10\,\mathrm{dB}$ in the complete absence of
coherence noise. The remaining $2\,\mathrm{dB}$ can be explained
by incomplete filtering (remaining coherent noise due to filter selection),
incoherent background noise being increased together with the sample
illumination, or by decreased signal due to the higher lateral sampling
frequency, where a lower modulation transfer of the camera is expected. 

\subsection{High imaging depth with off-axis holoscopy}

In FF-SS-OCT optimal lateral resolution is limited to a region of
the Rayleigh length around the focal plane. Due to the limited NA
of the optical system of the eye, this does not exceed the thickness
of the retina significantly. But imaging the anterior chamber or even
the whole eye by a single shot measurement with good total resolution
is still a challenge for OCT and needs ways to overcome the limited
depth of focus.

\begin{figure}[h]
\centering{}\subfloat[]{\includegraphics{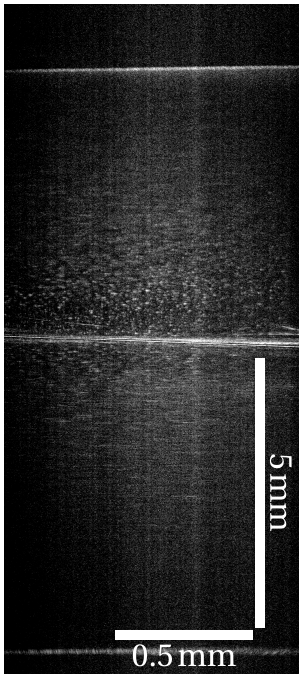}

}~\subfloat[]{\includegraphics{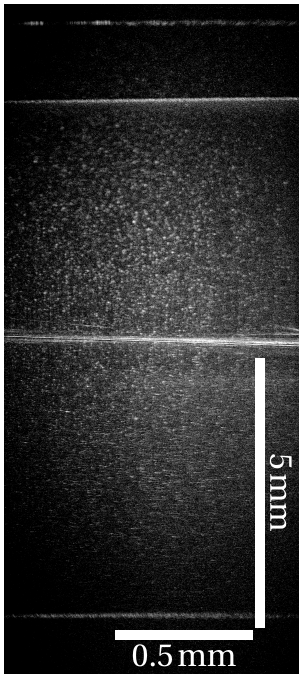}

}\caption{B-scan of nanoparticles randomly dispersed in polyurethane resin.
Due to full-range recording the B-scan covers a depth of about $10\,\mathrm{mm}$.
The central white structure marks the zero-delay line. At the imaging
NA of $0.07$ full lateral resolution was achieved only over a small
axial depth of approximately two Rayleigh lengths $2z_{R}\approx100\,\text{\textmu m}$
(a). When using digital refocusing techniques, the depth of focus
was increased to nearly the full imaging depth (b). \label{fig:NPL}}
\end{figure}

Holoscopy applies refocusing techniques to the full-field OCT signals,
restoring lateral resolution over tens of Rayleigh lengths. Combined
with an off-axis reference wave maximal use of the coherence length
of the swept laser source can be made, if reconstruction is only applied
to the signal term. Fig.~\ref{fig:NPL} shows a scattering sample
of polyurethane resin doped with iron-oxide nanoparticles of sizes
between $200\,\mathrm{nm}$ and $800\,\mathrm{nm}$ \cite{Woolliams2010},
where the signal term over positive and negative path lengths was
successfully refocused. By using a numerical reconstruction technique
for holoscopy \cite{Hillmann:11,Hillmann2:12} backscattered light
was detected with almost optimal lateral resolution in the full-range
images over a depth of more than $10\,\mathrm{mm}$.

\begin{figure}[h]
\begin{centering}
\includegraphics{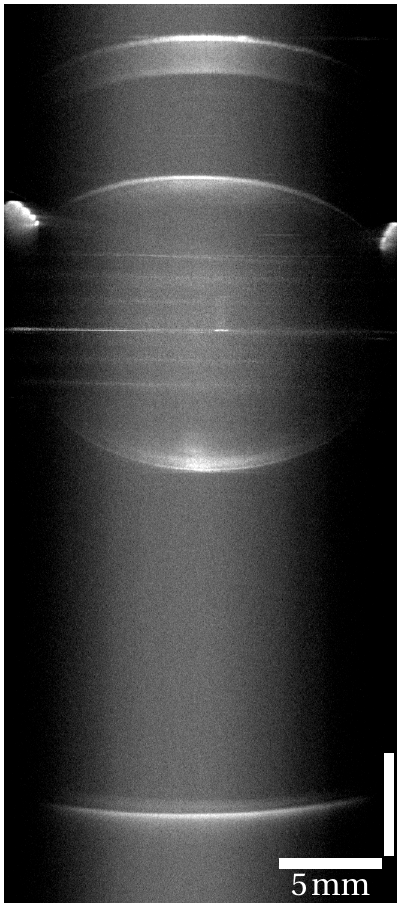}
\par\end{centering}
\caption{B-scan of an entire porcine eye including cornea, lens, and retina
with an entire measurement depth of more than $25\,\mathrm{mm}$.
During data acquisition the focus of the imaging optics was approximately
on the retina. Later the image was numerically refocused to the iris.
The focus on the retina could not be obtained as no lateral structures
were visible to ensure correct focusing. \label{fig:B-scan-of-an-porcine-eye}}
\end{figure}

Imaging the whole eye with FF-SS-OCT or holoscopy suffers from strong
reflections at the different optical interfaces of the eye, which
we expect to introduce significant autocorrelation noise or an increased
noise floor when being incoherent with respect to the sample light.
With off-axis recording and filtering of non-signal terms a complete\textit{
ex vivo} porcine eye, with both anterior and posterior segment, i.e.~retina,
lens, and cornea was imaged in a single data set (Fig.~\ref{fig:B-scan-of-an-porcine-eye}).
The sweep range of the light source was reduced to $20\,\mathrm{nm}$,
sacrificing resolution for a total measurement depth of more than
$25\,\mathrm{mm}$ when sampling the sweep with 2048 data points.
By numerical refocusing on the signal term both retina and cornea
were reconstructed sharply. To reduce noise, we acquired and averaged
9 volume data sets. Reflection induced artifacts and remaining DC
signals were identified by their constant phase values in all images
and removed numerically. 

\section{Conclusion and outlook}

We demonstrated full-range imaging and suppression of autocorrelation
artifacts by off-axis FF-SS-OCT and holoscopy. Dramatic improvement
of image quality was shown for samples with highly reflecting layers
such as a scotch tape roll. Furthermore, \textit{in-vivo} images of
the human retina show an increased SNR due to the reduction in coherence
noise. Parallelization of OCT imaging allows to increase the radiant
flux which is directed into the eye without damaging the retina; the
shown data was not yet limited by safety standards \cite{DIN:60825,ISO:15004}
and thus further increase in sensitivity is expected. Imaging speed
of 39~MHz equivalent A-scan rate on-axis and 7~MHz off-axis were
achieved, values which are two to three orders higher than todays
clinical devices. These values are limited by the camera speed and
the number of wavelengths sampled. By reducing resolution or imaging
depth even higher values are possible. Since scanning does not limit
the imaging speed multi-kHz volume rates are possible, which could
measure fast retinal changes with submicrometer axial resolution.
In the neuronal retina quality is comparable to scanned OCT system
considering the high imaging speed. However, choroidal structures
suffered from a strong background of multiply scattered light which
could not be reduced by off-axis imaging. The presented \textit{in-vivo
}images show the potential of parallelized imaging of the neuronal
retina, i.e., volumetric imaging at increased speed, resulting in
full phase stability which is hardly achieved by the fastest spectrometer
based OCT systems.

Holoscopy may also enable the imaging of the complete eye, i.e., anterior
and posterior segment, with high resolution. The filtering of the
image term in off-axis geometry doubles the imaging depth and suppresses
spurious autocorrelation signals from ocular surfaces. A porcine eye
was successfully imaged from cornea to retina using numerical refocusing
of the signal term. However, the coherence length of the light source
was not sufficient to acquire an entire human eye \textit{in vivo},
yet.

Although off-axis imaging only uses parts of the lateral Fourier-space
as oversampling is required to correctly capture the object wave field,
using an additional angle for the reference wave that is different
to the one used, a second signal could be acquired on the same camera.
This way two independent wave fields could be multiplexed in a single
image, which could for example be used for polarization sensitive
measurements. 

\section*{Acknowledgments}

This work was sponsored by the German Federal Ministry of Education
and Research (Program ``Innovative Imaging and Intervention'', contract
numbers 98729873C and 98729873E). 

\bibliographystyle{osajnl}
\bibliography{Holo,HoloPapers2,Bessel,FullRange,Books,OCT,NFFT,DigitalHolography,OCM,Microscopy,Intro,Other}

\end{document}